# 3D Mapping of Static Magnetic Field Magnitude and Axial-Components around a total body 3T MRI clinical scanner

**Francesco Girardello[1], Maria Antonietta D'Avanzo[2], Massimo Mattozzi[2], Victorian Michele Ferro[3], Giuseppe Acri[3,\*], Valentina Hartwig[1]**

[1] Institute of Clinical Physiology, CNR, Via G. Moruzzi 1, 56124 Pisa, Italy

[2] Department of Occupational and Environmental Medicine, INAIL, Via Fontana Candida 1, 00078, Rome, Italy; m.davanzo@inail.it (M.A.D.); m.mattozzi@inail.it (M.M.)

[3] Department of Biomedical, Dental, and Image Sciences, University of Messina, Via Consolare Valeria, 98125 Messina, Italy

\* **Correspondence:**
Corresponding Author
giuseppe.acri@unime.it (G.A.); valentina.hartwig@cnr.it (V.H.)

**Abstract**

*Objective*. The technology employed in magnetic resonance imaging (MRI) systems has evolved continuously, resulting in MRI scanners with stronger static magnetic fields (SMF) B0, faster and stronger gradient magnetic fields, and more powerful radiofrequency transmission coils. The most well-known hazard associated with an MRI environment is the projectile effect due to Spatial Field Gradient (SFG). Furthermore, movement through the SFG generates a time-varying magnetic field, which in turn induces a voltage in body tissues. This has the potential to result in a range of physiological symptoms, including headache, nausea, vertigo, phosphenes, numbness, tingling, loss of proprioception, and balance disturbances.

*Approach*. The methodology outlined in this study provides a comprehensive and reliable approach to creating a 3D map of the SMF (fringe field) around a clinical MRI facility. The methodology involves measuring the unperturbed B field, including magnitude and axial components, in specific points and subsequently performing a mathematical procedure involving fitting and interpolation.

*Main results*. Fringe field magnitude and axial components 3D maps are presented for a 3T whole-body MRI scanner for clinical application located in a hospital facility.

*Significance*. The map obtained could be used for a number of purposes, including the evaluation of hazard. This could be achieved by using digital tools to create a simulation of all types of MRI workers movements within the facility. The map could also be used for the training and education of MRI operators, with a view to establishing best practices. The estimation of magnetic field axial components represents a valuable enhancement, as these data can be used to calculate induced electric fields during rotational movements, such as those of the head or torso.

## 1 Introduction

Magnetic resonance imaging (MRI) is a widely utilised tool in both medical research and diagnostic imaging. In contradistinction to ionising radiation, MRI employs electromagnetic radiation, which is characterised by an energy level that is insufficient to dislodge electrons from atoms or molecules.

Nevertheless, the principal hazard linked with MRI is posed by the static magnetic field (SMF), which, for the majority of clinical scanners, is constantly active (Greenberg et al., 2020).

The technology employed in MRI systems has evolved continuously, resulting in MRI scanner with stronger SMF (B0), faster and stronger gradient magnetic fields, and more powerful radiofrequency transmission coils (Fagan et al., 2021). Many safety investigations have been carried out on 1.5T scanners, although in the last few years, many centres have installed magnets of 3.0T and above. It is imperative that all personnel involved are cognizant of and adept in the identification and mitigation of MRI hazards. Some of these hazards may include projectile accidents, whereby the powerful magnetic field produced by the MRI machine can cause metallic objects to fly into the air and possibly hurt medical professionals (Kim and Kim, 2017; Hartwig et al., 2018, 2021).

Spatial Field Gradient (SFG), is defined as the rate of change in the magnetic field as a function of position around the MRI system. The SFG is known to decrease with increasing distance from the extremities of a standard cylindrical, horizontal-field magnet and it is responsible for the attractive force on ferromagnetic objects. The SFG characterises the temporally fixed spatial gradient magnetic field surrounding the MR system, with its regional value depending on B0 and scanner shielding (McRobbie et al., 2017). Passive implanted items, including vascular clips and protheses, and active implanted medical devices (AIMDs), such as pacemakers and cochlear implants, are also susceptible to forces and torques in the MRI environment, which can result in significant impairment (Zradziński et al., 2018; Mattei et al., 2019; Steckner et al., 2024). Consequently, AIMDs and all other medical equipment intended for use in the MRI environment are typically excluded from the 0.5 mT (5 G) fringe field. It is customary for MRI scanner manufacturers to provide a map or chart of the SMF magnitude around the scanner, also referred to as fringe field or stray field, typically in the form of an isogauss lines map, which indicates the strength of the field at specific locations. These are utilized by MRI personnel to ascertain whether the maximum field to which an implant will be subjected exceeds the "MR Conditional" value indicated on the label (Simmons et al., 2016; McRobbie, 2020). As the information is presented in different ways by each manufacturer, it is important that users understand how to interpret it for the purposes of the relevant scanner.

Furthermore, in MRI workplaces, movement through the SFG acts as a time-varying magnetic field (motion-induced TvMF) (ICNIRP, 2014), inducing a voltage in electrically conductive materials, such as biological tissues, in accordance with Faraday's law (McRobbie, 2020). Consequently, rapid body movements generate a substantial electric field within the tissue, potentially resulting in a range of physiological symptoms, including headache, nausea, vertigo, phosphenes, numbness, tingling, loss of proprioception, and balance disturbances (König et al., 2024). Despite the extensive literature on patient safety in MRI treatments, it is crucial to acknowledge the significant hazards faced by medical workers involved in these procedures. A variety of scientific publications exist on the occurrence of short-term sensory effects as well as on the occurrence of neurocognitive and neurobehavioural effects. However, this disturbance is typically transient. Long-term effects may include a predisposition for hypertension and sleep disturbances. The data concerning the exposure of healthcare workers to magnetic fields during pregnancy has been determined that there are no particular deviations with regard to the duration of pregnancy, premature births, miscarriages, and birth weight. The paucity of epidemiological studies in this area is a key concern: there is a considerable need for high-quality data, particularly on the consequences of long-term exposure to electromagnetic fields from clinical MRI (König et al., 2024).

From a more technical standpoint, a number of literature studies regards the risk assessment for workers exposure to SMF and motion-induced TvMF in MRI environment. The primary objective of the exposure assessment is to verify compliance with the established exposure limits stipulated in the current regulatory framework, such as the directive issued by the European Parliament and the Council of the European Union (European Parliament and Council of the European Union, 2013) and the guidelines of the International Commission on Non-Ionizing Radiation Protection (ICNIRP) (ICNIRP, 2009, 2010, 2014). Additionally, it is intended to characterise potential exposure scenarios within the context of epidemiological studies on MRI workers exposure. This assessment is identified as a high priority to address the existing knowledge gap concerning the associated health implications. As previously stated, the unperturbed field in the MRI environment can be determined by using the isogauss line maps provided by the scanner manufacturers (Karpowicz et al., 2007; Betta et al., 2012). However, given that most of the extant maps do not provide a comprehensive representation of the magnetic fields near magnets, the utilisation of isogauss line maps constitutes a rough method by which to evaluate workers' exposure.

The magnitude of a SMF can be directly measured using commercial survey meters (Andreuccetti et al., 2017; Gurrera et al., 2019; Hartwig et al., 2022). However, this method is capable of providing a limited number of values for magnetic field in specific locations within the MRI room. Consequently, this approach can only provide an approximate indication of the fringe field.

A substantial number of studies have been documented in the extant literature, the objective of which was to evaluate the exposure of MRI personnel to SMFs and motion-induced TvMF in a spatially heterogeneous magnetic field (Acri et al., 2015, 2018; Sannino et al., 2017; Belguerras et al., 2018). The majority of these studies were based on theoretical models or personal measurements of exposure to magnetic fields, using dosimeters. A digital tool has been presented in (Hartwig et al., 2011) and (Hartwig et al., 2014) that simulates the linear path followed by an MRI worker during a routine procedure and calculates the induced electric field in a simple model of the body. The tool utilised the distribution of the fringe field, derived from the isogauss line map of a particular MRI scanner. Subsequent studies by Gurrera et al. (Gurrera et al., 2019) presented an analytical model based on the map of the stray field of a magnetic dipole as approximation of the magnetic field straying from a closed full-body MRI scanner. Later, they added to the model an accurate analysis of human movements: whole-body movements were recorded in a gait laboratory set up to reproduce the workspace of a room with a whole-body MRI scanner, using a stereophotogrammetric system to obtain the speed trend during the movements (Gurrera et al., 2021). In this study, the stray magnetic field surrounding the MRI scanner is approximated using a simple dipole model, which disregards the specific architecture and shielding of individual machines.

In this study, we present a methodology for creating a comprehensive and reliable 3D map of the fringe field of a general clinical MRI facility. The methodology employed in this study involves the measurement of the unperturbed magnetic field B in specific points, followed by a mathematical procedure involving fitting and interpolation. This process is then utilised to obtain the B values throughout the entire room, with a resolution of 1 × 1 × 1 cm. Unlike previous similar studies, here the map of the magnitude of B (|B|) as well as each of its axial components (Bx, By, Bz) is estimated.

The map that has been obtained could be utilised for a number of purposes including the evaluation of hazard using digital tools to create a simulation of all types of MRI workers movements within the facility, as well as for the training and education of MRI operators with a view to establishing best practices.

## 2    Materials and methods

### 2.1    Measurement Acquisition

Measurements of the magnetic field and its components in the MRI rooms were carried out using a commercial HP-01 magnetometer field analyzer (Narda Safety Test Solutions, Savona, Italy) in the area where workers typically move during their daily activities. The instrument provides a resolution of 100 nT for field strengths up to 50 mT, and 100 µT for values above 50 mT. The manufacturer specifies a DC measurement accuracy of 1% ($\sigma_{exp}$). This level of precision is suitable for evaluating fringe fields in MRI environments, particularly when considering regulatory thresholds such as 0.5 mT and 3 mT, which are commonly used to define controlled access zones. The spatial sampling density and measurement protocol were designed to capture magnetic field variations relevant to personnel movement and to reliably detect threshold crossings.

Static magnetic field measurements were taken on a 10 × 10 cm grid on planes parallel to the ground (xz planes) at different heights from the floor, in the area of interest (near the scanner gantry).

The measurements were taken at three different heights: y = 95 cm, y = 138 cm, and y = 160 cm. For a 1.70 m tall worker, these heights correspond to the genitals, torso (heart), and the head, respectively.

The cover area was the frontal area to the left of the patient's bed starting from the MRI gantry and extending up to 115 cm along the z-axis (parallel to the patient's bed) and 75 cm along the x-axis (perpendicular to the patient's bed). The measurements were acquired following a procedure similar to the one described in (Hartwig et al., 2024). A total of 252 measurements were performed. The plane on which the measurements were taken is schematically represented by the red square in **Figure 1**.

## 2.2 Data Processing on a Single Plane

All calculations described in this work were performed with homemade MATLAB®, R2020b (MathWorks, Inc., Natick, MA, USA) scripts.

As the first step in data processing, we performed quality control to eliminate potential outliers from the actual measurements. This procedure proved essential considering the extremely sensitive nature of the readings taken with the gaussmeter within the experimental environment. The precision of sensor positioning is indeed critical, as even millimetre deviations from the target position can generate significant alterations in the measured values. The identification of outliers was therefore conducted following an approach based on the expected trend of values along the x and z axes, thus ensuring the integrity and reliability of the dataset used in subsequent analysis phases.

Subsequently, to create a comprehensive magnetic field map of the entire MRI room, it was necessary to extend the measurements from each plane across the entire cross-section of the room. To model the spatial distribution of the magnetic field, we implemented a parametric fitting approach using various families of non-linear functions.

The need for multiple function families was dictated by the complexity of the analysis, which required independent fitting along both the x-axis and z-axis for each of the three magnetic field components and for its magnitude. The fitting functions ranged from simple exponential combinations to more complex forms incorporating polynomial terms modulated by exponential decay.

The complexity of the functions used was progressively scaled, with models employing from 4 to 6 free parameters, in order to adequately capture both the short-range and long-range behaviour of the measured magnetic field. Specifically, we explored functions that combine pure exponential terms, products of exponentials and polynomials, and hybrid forms with both linear and quadratic exponential dependencies in the spatial variable, optimizing the choice of the fitting function based on the specific characteristics of the trend to be modelled in each measurement region.

The choice of the optimal fitting functions for the experimental data was carried out through a systematic approach based on the comparison of the reduced chi-squared value. For each function the reduced chi-squared $\chi^2_{red}$ was computed according to the follows equation:

$$\chi^2_{red} = \frac{\sum \left(\frac{residual_i}{B_i}\right)^2}{n - p}$$

where residual represents the difference between the measured value and the model prediction for the i-th point, B is the measured value (|B| or Bx, By, Bz), n is the total number of experimental points, and p is the number of parameters in the fitting function. The closer the value is to 1, the better the functions approximate the experimental trend.

The fitting uncertainty (σfit) was calculated as the residual standard deviation - the standard deviation of the residuals between observed and predicted values - using the formula provided in standard regression methodology. This approach aligns with established statistical practices for assessing the goodness of fit and quantifying model-related uncertainty (Dodge, 2008).

Through the previously described fitting procedures, it was possible to extend the analysis of the magnetic field to the peripheral regions (shown in blue in **Figure 1**) of quadrant 1 of the room. To complete the mapping of this area, an additional interpolation phase was necessary in the outermost region (highlighted in green). A function was then used to interpolate the scattered data with a natural neighbour interpolation method. This specific interpolation technique was selected following a comparative analysis with other available methods, proving to be optimal in terms of continuity and physical consistency in the magnetic field reconstruction within the region of interest (Hartwig et al., 2024). For interpolated data points, the uncertainty was evaluated based on established principles of error propagation in spatial interpolation procedures (White and Saunders, 2007). The interpolation uncertainty was determined as:

$$\sigma_{interp} = \sigma_{method} + \beta \times d_{min}$$

where $\sigma_{method}$ represents the inherent uncertainty that depends on the type of interpolation used, $\beta$ is the spatial distance factor (0.05% per cm), and $d_{min}$ is the nearest distance to measured data points.

The intrinsic symmetries of the magnetic field can be exploited to achieve a complete mapping of the entire space. B values are typically symmetrical horizontally (around the central x-axis), vertically (around the central y-axis) and radially (around the central z-axis). For quadrant 2, symmetry with respect to the x-axis was applied: the values of the magnetic field magnitude and the By and Bz components were mirrored, while the Bx component was inverted in sign. Similarly, using symmetry with respect to the z-axis, the mapping was extended to quadrant 4, this time inverting the sign of the Bz component.

In quadrant 3, a more sophisticated interpolation procedure was necessary, considering the presence of a spherical region with a diameter of 60 centimetres, situated at the isocentre of the magnet.

Within this region, the magnitude of the magnetic field remains constant at 3 Tesla. The procedure involved the preliminary insertion of a disk - whose size is dependent on the height of the reference plane - representing the intersection of the isocenter sphere with the analysis plane. The interpolation process for the magnitude in quadrant 3 was carried out in two stages:

1. An initial interpolation matching the resolution of the experimental data (10 cm), using a natural neighbor interpolation method to generate a regular grid of points;
2. A subsequent high-resolution interpolation (1 cm) for 2-D gridded data in meshgrid format, using the cubic method.

This procedure allowed us to obtain, for each xz plane, a complete and high-resolution mapping of the magnetic field magnitude.

### 2.3 3D Mapping of the MRI Room

The creation of a three-dimensional map of the fringe field in the MRI room necessitates the acquisition and interpolation of a minimum of three complete planes, as previously outlined. The accuracy of three-dimensional interpolation process is directly proportional to the number of planes acquired. In order to optimise the volumetric reconstruction, the symmetry of the magnetic field with respect to the y-axis is exploited, thereby effectively doubling the number of planes available for interpolation. In this process, the By component of the magnetic field is appropriately sign-inverted, and the symmetrisation is performed with respect to the y-axis.

The implementation of the software has been structured according to a modular approach, which is divided into two distinct phases. The initial phase is dedicated to the management of the size data of the designated MRI room, which encompasses both the characteristic parameters of the room itself and the interpolated planes. The subsequent phase involves a general three-dimensional interpolation function that, starting from the previous data, generates the complete volumetric map. The comprehensive uncertainty was obtained through quadratic combination of all uncertainty sources:

$$\sigma_{\text{tot}} = \sqrt{\sigma_{\text{exp}}^2 + \sigma_{\text{fit}}^2 + \sigma_{\text{interp}}^2}$$

Each individual uncertainty term has been discussed in detail in the preceding sections. Specifically, $\sigma_{exp}$ represents the experimental uncertainty associated with the measurement instrument, $\sigma_{fit}$ refers to the uncertainty introduced by the fitting procedure, and $\sigma_{interp}$ accounts for all interpolation procedures performed, including both 2D plane interpolations and subsequent 3D spatial interpolation. This systematic uncertainty evaluation provides reliable error quantification across the entire mapped region, acknowledging that interpolation uncertainty grows with increasing distance from experimental measurement locations.

For occupational exposure assessment, a total relative uncertainty of up to 10 % was considered acceptable, in line with common practice for exposure evaluation in realistic workplace scenarios. Such levels of uncertainty are sufficient to reliably identify exposure thresholds and classify controlled-access zones without compromising safety-related decision-making accuracy (Mild et al., 2009).

This software architecture allows for flexible and scalable management of different MRI rooms, requiring only the specification of the reference room to generate the corresponding complete mapping. The 3D interpolation process initially involves creating a matrix containing the interpolated planes, followed by the application of volumetric interpolation using a linear interpolation method. This approach produces a 3D map of the entire room for the magnetic field magnitude, while for the individual components, it generates 3D maps of the areas in front of and behind the magnetic resonance machine. The described interpolation methodology is generalizable to other MRI rooms, provided that the experimental measurements cover a minimum surface of 70 cm × 70 cm along the x and z axes. The acquisition planes can be positioned at arbitrary heights. The presented methodology has been tested for two hospital facilities (FTGM Ospedale del Cuore - Massa and FTGM Ospedale San Cataldo – Pisa), each of which is equipped with an MRI scanner with a B0 of 3 T from two different manufacturers.

## 3    Results

The results presented in this section pertain to the implementation of the aforementioned procedure on a 3T total body MRI scanner for clinical application, situated at FTGM Ospedale del Cuore – Massa, Italy. As an example, **Figure 2** shows the fitting procedure for the data in terms of |B|, Bx, By, and Bz values along the x-axis for a fixed z = 30 cm from the MRI bore and y = 138 cm from the floor. In this particular instance, the estimated chi-squared value is equivalent to 1.174 for the magnitude, while for the components we have Bx = 1.117, Bz = 0.605, and By = 1.498.

In **Figure 3**, the outcomes of the fitting procedure and the initial interpolation step for the magnitude |B| are presented. The ensuing results are presented in conjunction with the axial components Bx, By, and Bz for the xz plane (y = 138 cm). **Figure 4** presents the results of a simulation of magnetic field lines (isogauss lines) for the three reference planes xz, located at y = 95, 138, and 160 cm. In **Figure 5**, the three axial views of the isogauss lines for the orthogonal planes through the isocentre (@ y = 105 cm from the floor) of the scanner are presented. This is the representation for the isogauss lines map that is typically provided by the scanner manufacturer.

Finally, **Figure 6** shows the two-dimensional map of the magnetic field magnitude, denoted by Btot, throughout the designated space. This representation accounts for the presence of the spherical area (30 cm radius) in which |B| is homogeneous and equal to 3 Tesla, centred in the isocentre of the MRI system.

The axial components of the magnetic field are illustrated in **Figure 7**, with reference to the plane positioned at y = 95 cm. In contrast to the field magnitude, the components are not defined in the central part of the room, since no information is available about the magnetic field components value inside the scanner.

Once consistent results have been obtained for all the planes considered, spatial interpolation is performed to reconstruct the three-dimensional map of the magnetic field within the room. In **Figure 8**, the magnitude of the magnetic field |B| is plotted within a cubic volume, with its centre situated at the isocentre of the MRI scanner. **Figure 9** presents a similar representation with the planes intersecting the isocentre. Finally, **Figure 10** provides a visualisation of the isogauss surface within the MRI room.

To evaluate the uncertainty associated with the three-dimensional magnetic field map, four representative points were selected in regions characterised by distinct data processing steps: direct measurement, mathematical fitting, two-dimensional interpolation, and full three-dimensional interpolation. The total uncertainty at each point was computed by combining the instrumental error with the specific contribution of each processing stage. The analysis revealed that uncertainty increases progressively from directly measured points to fully interpolated regions, with the highest uncertainty value of 5.58% found in areas requiring complete three-dimensional interpolation.

## 4    Discussion

The safety of MR scanners is evolving with the technology and how they are used. MRI scanners with stronger SMFs (B0), faster and stronger gradient magnetic fields, and more powerful radiofrequency transmission coils are increasingly common in clinical and research settings. It is imperative that all MRI workers be aware of the potential hazards and are fully informed of the safety procedures that should be followed. Furthermore, it is essential for operators who move within the MRI room on a daily basis to be fully mindful of the fringe field present in their specific work environment. Finally, in order to proceed with the estimation of exposure to static and motion-induced TvMF, using digital simulation tools, it is first necessary to accurately estimate the 3D magnetic field distribution around a specific scanner.

It has been observed that isogauss line plans provided by manufacturers, often fail to provide a sufficiently detailed map of the fringe field. Specifically, the zone closest to the gantry, which is the one with the highest spatial gradient value, is not detailed enough (Hartwig et al., 2021, 2024). Moreover, the isogauss line plans given by the manufacturers are generally related to the static fields in the absence of additional shielding (Capstick et al., 2008).

Furthermore, the information is usually limited to the magnitude of the fringe field, with its axial components not being observed. As the fringe field is a vector quantity, it is necessary to consider Bx, By, and Bz in order to implement Faraday's law in its complete form. It is imperative for the estimation of exposure in relation to electric fields induced by complex motions, such as rotation or torsion (ICNIRP, 2014; Gurrera et al., 2019).

Another potential methodology for obtaining a 3D map of the fringe field involves the use of mathematical modelling tools, such as simple dipole approximations (Gurrera et al., 2019), as previously explored in the literature. However, these models do not consider scanner-specific factors, such as active or passive shielding systems, the construction features of the installation room or the presence of nearby equipment. Consequently, 3D maps derived from these simplified models may misrepresent the actual fringe field distribution, often overestimating the field strength due to the absence of field-reducing elements. This is a particular limitation of newer MRI systems, which employ advanced active shielding technologies designed to reduce the field outside the magnet coils. These systems reduce the extent of the fringe field and generate steeper magnetic field gradients near the gantry. For all magnet types, additional passive shielding, such as strategically placed iron or high-permeability steel plates, also contributes to field shaping (McRobbie et al., 2017).

In this work, we propose a comprehensive methodology for reconstructing the three-dimensional map of the SMF (fringe field) around an MRI machine. The objective of the proposed approach is not to verify compliance with imposed safety limits, but rather to develop advanced educational tools for personnel operating in MRI environments. These tools will help personnel to identify and avoid higher-risk conditions and establish best operational practices.

Although the highest exposure relevance is typically associated with the zone closest to the scanner, that we identified with a red box in Figure 1, our decision to extend the field mapping beyond this region was motivated by practical and methodological considerations. In clinical environments, personnel or equipment may occasionally operate near the red-zone boundaries, where magnetic fields—though below critical thresholds—can still pose hazards, especially in the presence of active implants or ferromagnetic materials. Moreover, capturing a wider spatial distribution of the magnetic field enables a more complete characterization of field spatial gradients, which is essential for future modeling of motion-induced electric fields and exposure trajectories. Therefore, the extended mapping provides both safety-relevant and technically valuable information for more comprehensive exposure evaluations.

In this context, the following key aspects emerge from the presented work:

1. A map of the fringe field that is both detailed and reliable can be obtained through specific measurements of B in the MRI room. This is followed by mathematical fitting and interpolation procedures.
2. The methodology described herein enables the determination not only of the magnetic field magnitude but also of its individual axial components, which are essential for evaluating exposure during rotational movements of the head or torso.
3. The developed software is characterised by a modular approach, which facilitates flexible and scalable management of different MRI rooms. This approach allows for the consideration of the particular characteristics and possible complexities of the specific environment and scanner. The methodology presented has been successfully tested in two hospital facilities with 3T MRI scanners from different manufacturers, thus demonstrating the general applicability of the approach to various clinical MRI environments.
4. The map obtained can be used for a number of purposes, including risk assessment, the simulation of workers movements within the facility, and operator training. Future developments of this work will involve the use of the 3D map of the magnitude and the axial components of the fringe field to estimate the exposure of parts of the human body, such as the head or torso, during complex movements including linear and vertical displacements, rotations and flexions in a specific MRI environment.

In conclusions, the spatially detailed SMF maps produced in this study provide a solid foundation for future investigations aimed at evaluating workers exposure during standardized complex movements, such as displacement, rotation, and flexion. This direction, already supported in the literature through exposimetric studies on moving MRI workers, represents a natural and relevant extension of our current work toward more realistic, trajectory-based exposure assessments.

## 5      Conflict of Interest

The authors declare that the research was conducted in the absence of any commercial or financial relationships that could be construed as a potential conflict of interest.

## 6      Author Contributions

FG, VH: Conceptualization, Data curation, Investigation, Methodology, Software, Visualization,

Writing – original draft, Writing – review and editing. VMF: Data curation, Writing – original draft.
MM, MDA: Project administration, Resources, Writing – review and editing VH, GA: Conceptualization, Formal Analysis, Funding acquisition, Project administration, Supervision, Writing – review and editing.

# 7 Funding

This research was funded by INAIL (National Institute for Insurance against Accidents at Work), grant number Bric 2022 CUP: J43C22001390005.

## 9  Figures captions

**Figure 1**. MRI room axial view. Yellow circle: 3T isocentre area; Red box: measurement area; Blue boxes: fitting area; Green box: interpolation area.

**Figure 2**. Example of a magnetic field fit obtained by the measured points throughout the room. The fit is along the x-axis with z fixed at 30 cm for the plane at y = 138 cm.

**Figure 3**. Results of the fitting and interpolation procedures in the right-anterior area with respect to the MRI gantry (@ y = 138 cm).

**Figure 4**. Simulated magnetic field lines (isogauss lines) reconstructed from the magnetic field data for the three reference planes at y = 95, 138, and 160 cm.

**Figure 5**. Three axial views of the isogauss lines for the orthogonal planes through the isocentre of the scanner (@ y = 105 cm from the floor).

**Figure 6**.  Calculated magnetic field magnitude |B| for the three reference planes xz.

**Figure 7**. Calculated magnetic field axial components for the plane xz at y = 95 cm.

**Figure 8**. 3D map of the magnetic field magnitude |B| in the area relative to positive x, y, and z axis (isocentre is at x = 0, z = 0, y = 105 cm).

**Figure 9**. 3D map of the magnetic field magnitude |B| in the entire MRI room (isocentre is at x = 0, z = 0, y = 105 cm). The external part has been removed to show the isocentre zone.

**Figure 10**. 3D Isogauss Surfaces.

**Figure 1**.

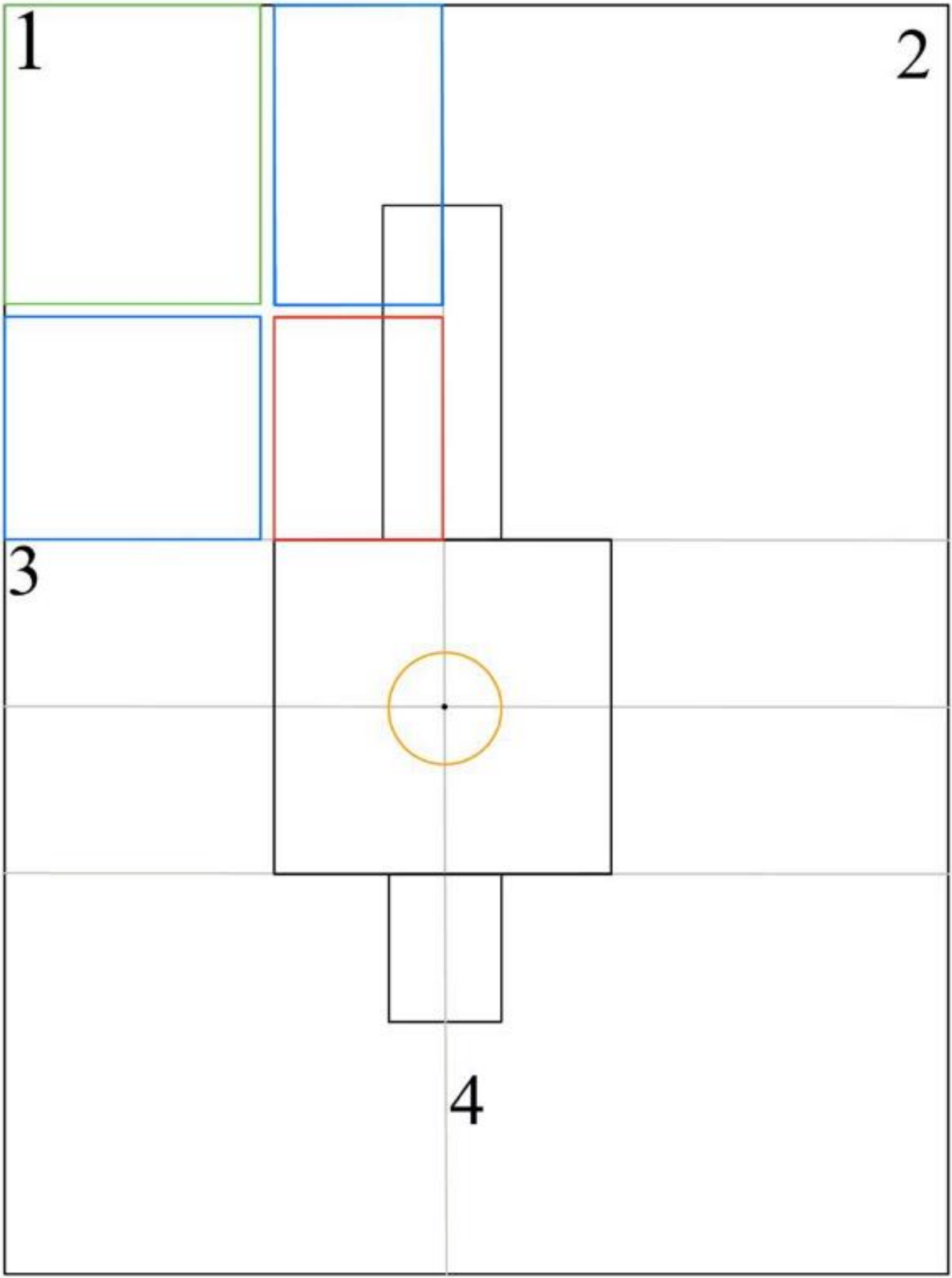

**Figure 2**.

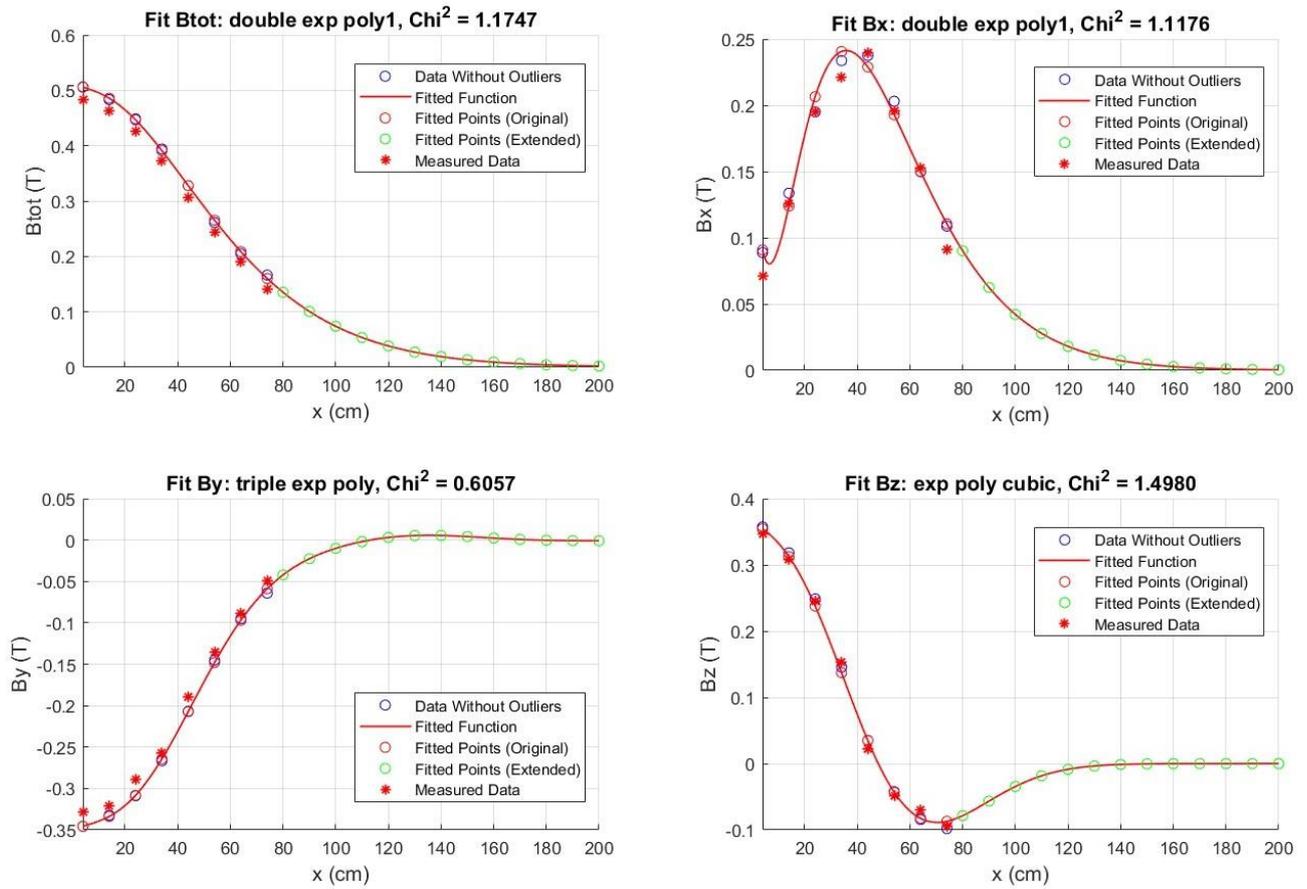

**Figure 3**.

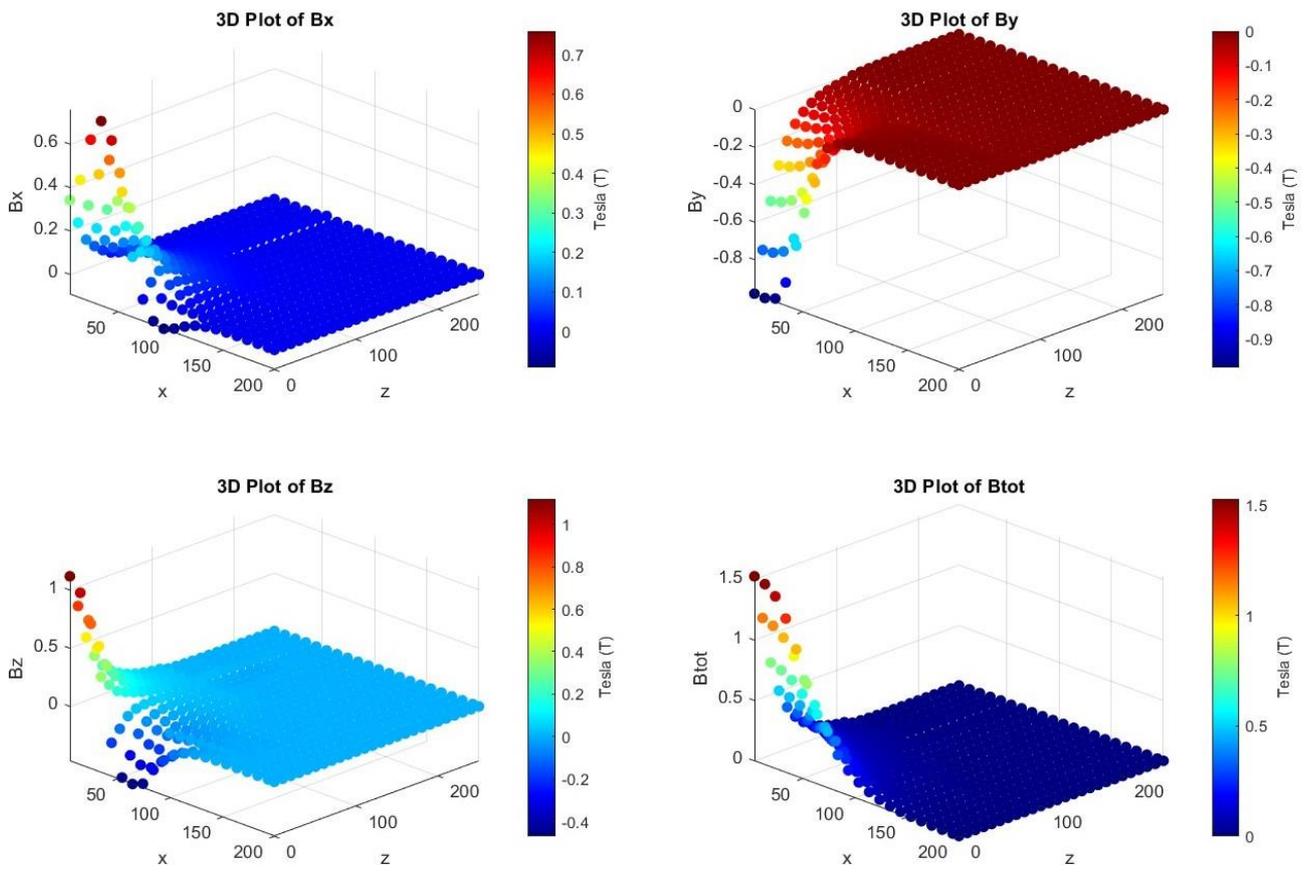

**Figure 4**.

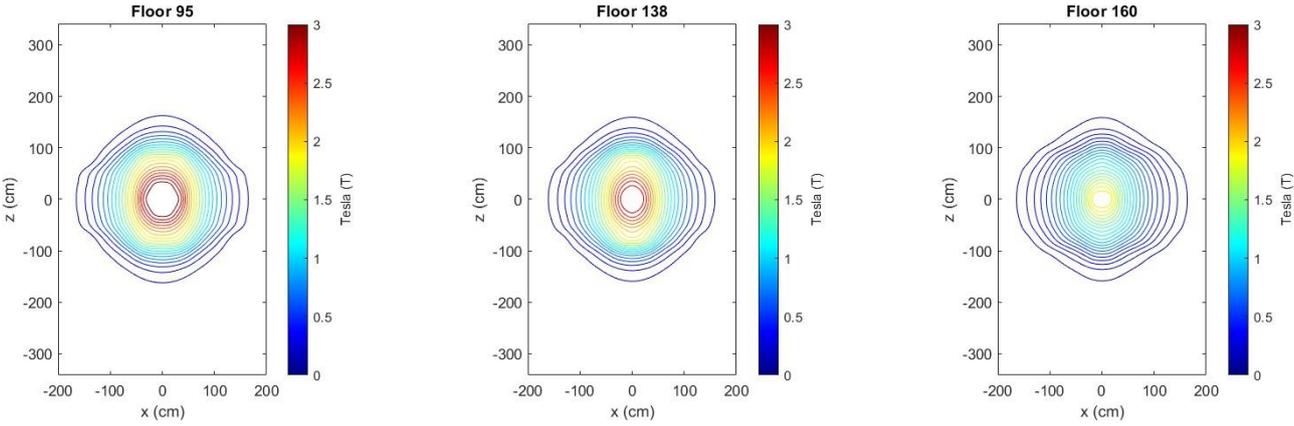

**Figure 5**.

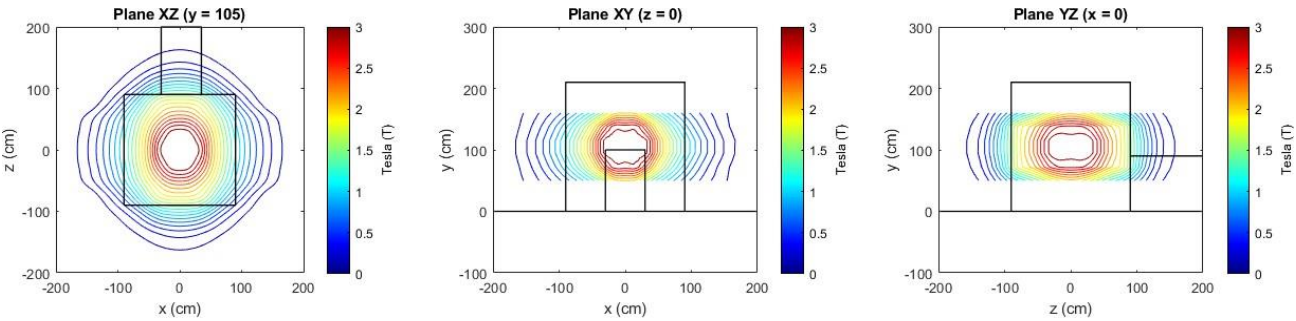

**Figure 6**.

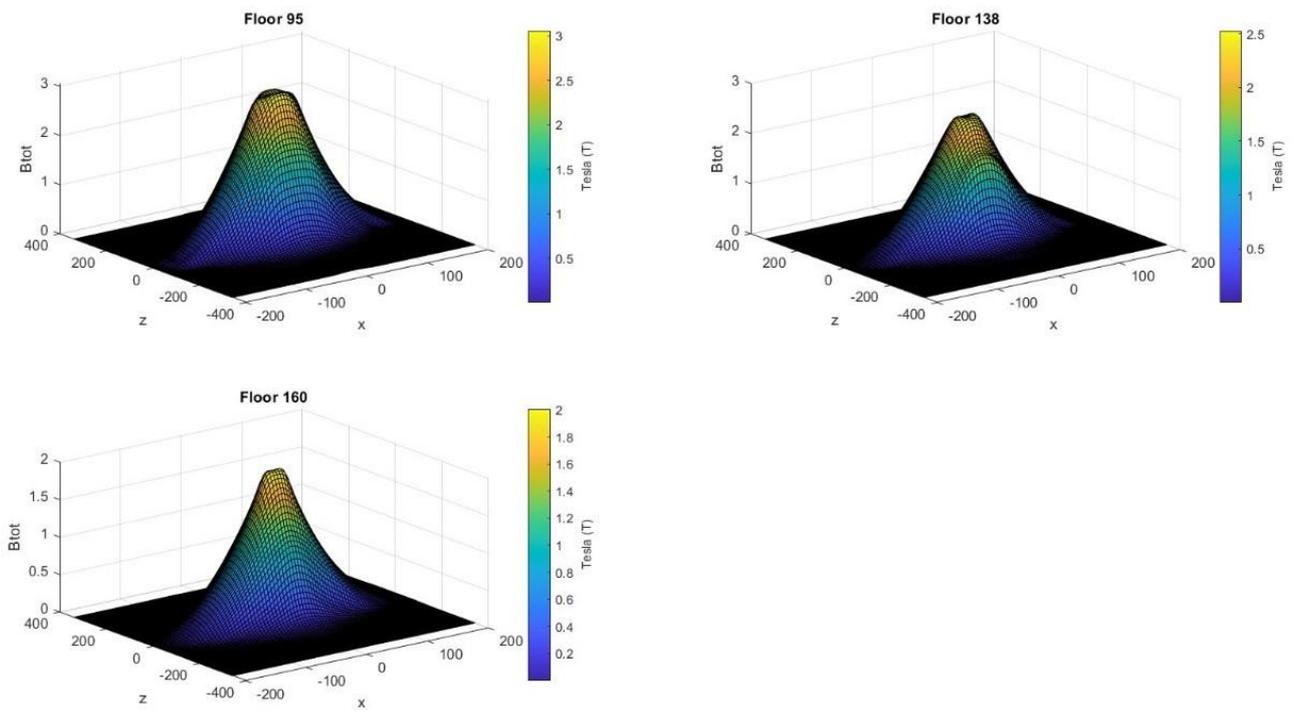

**Figure 7**.

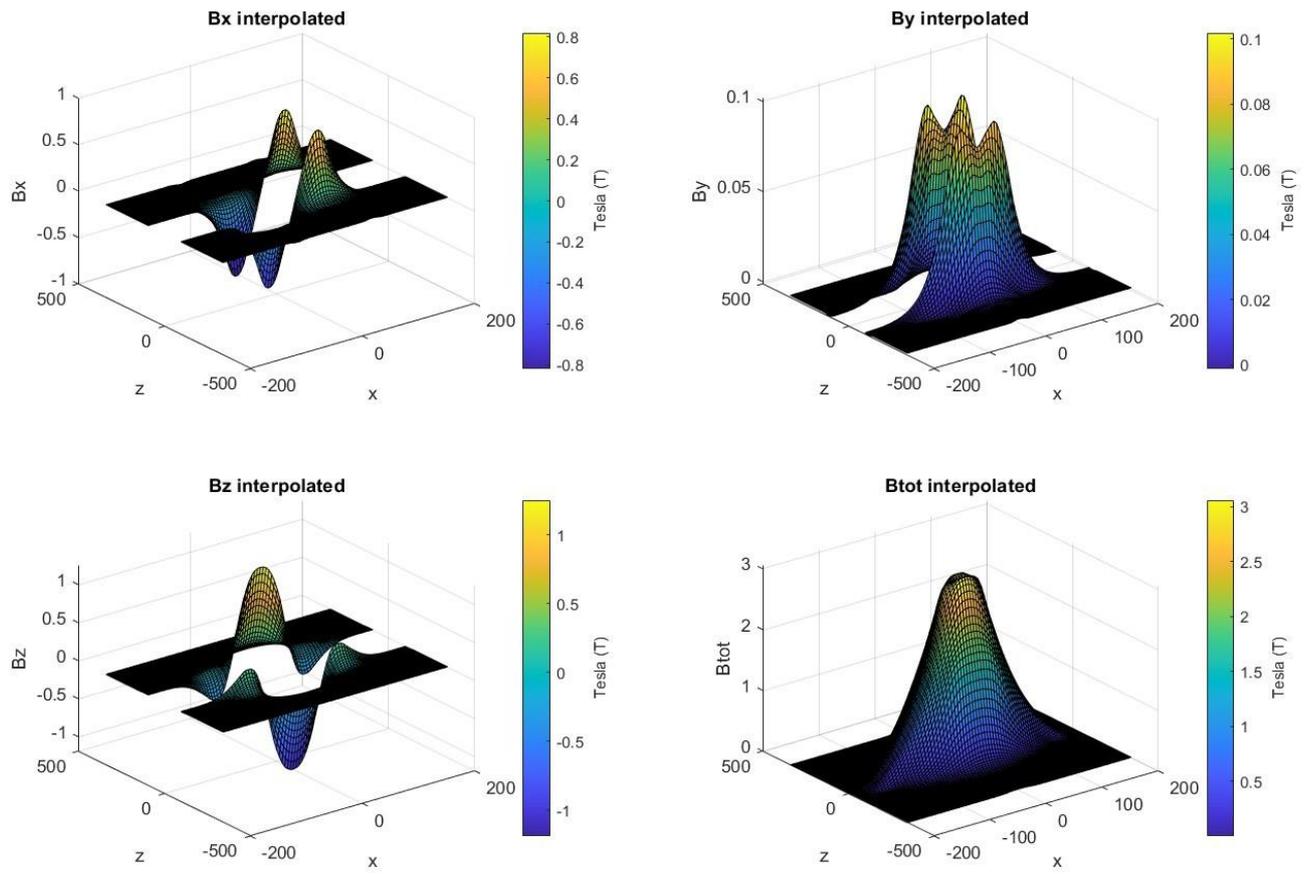

**Figure 8**.

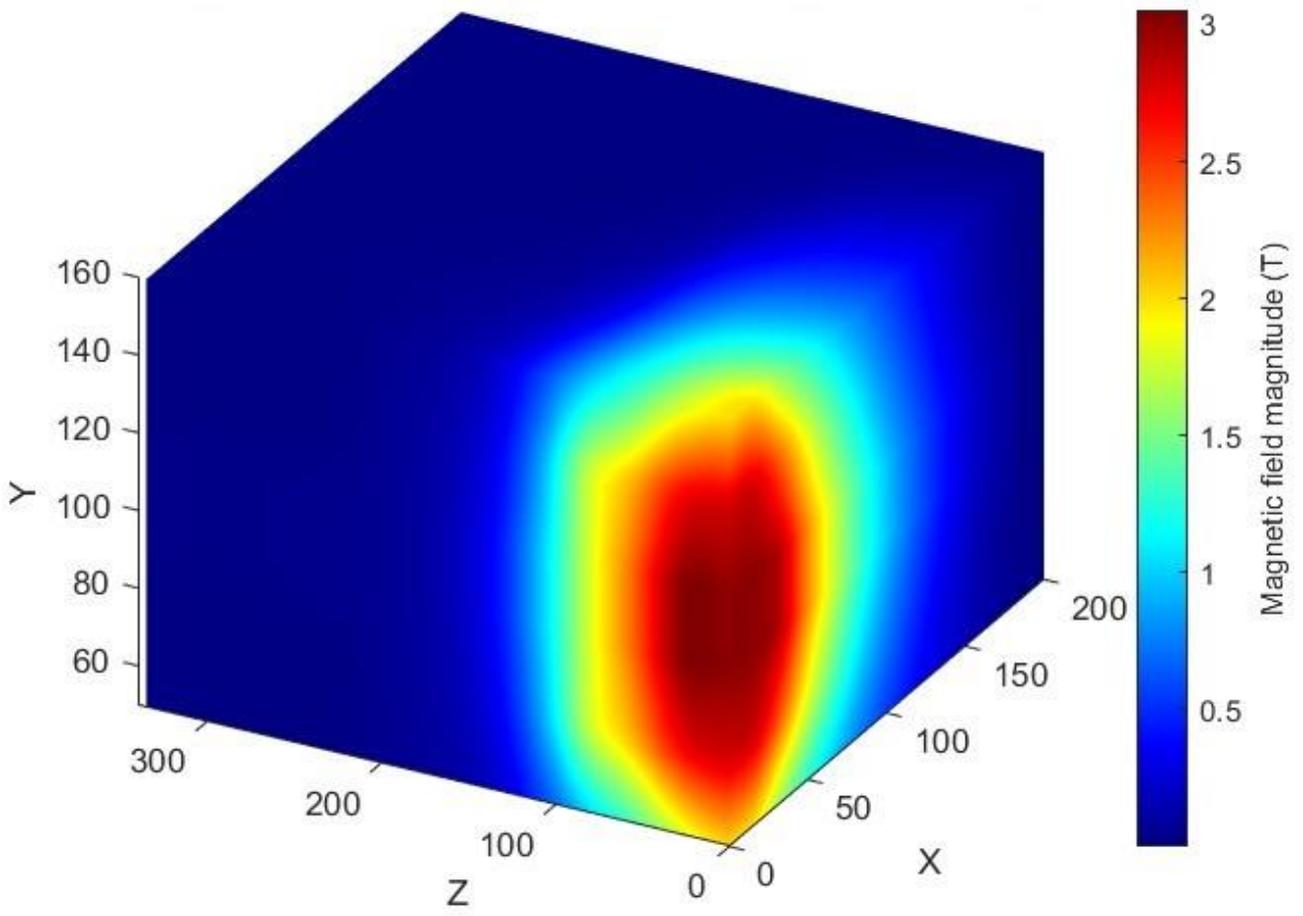

**Figure 9**.

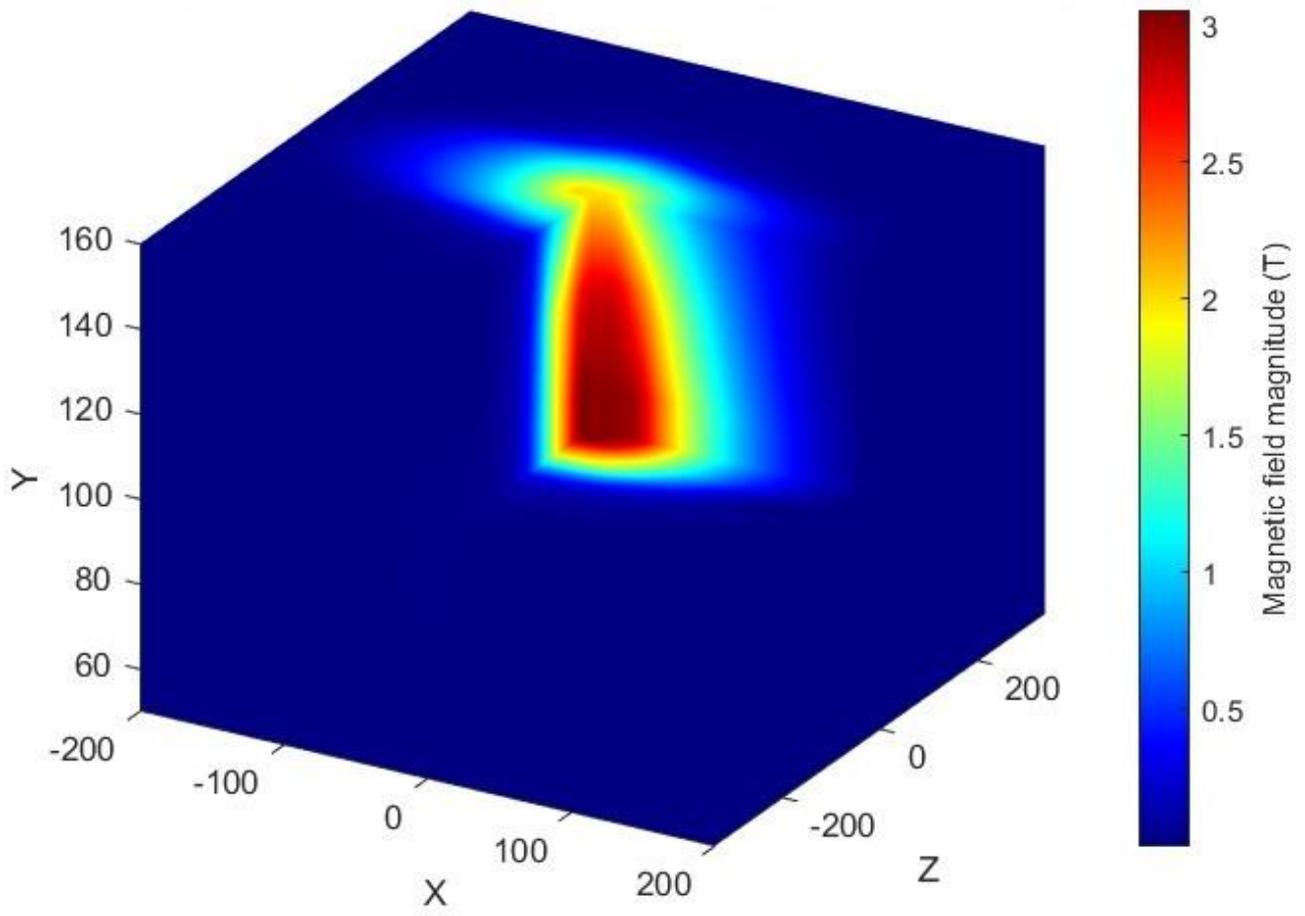

**Figure 10**.

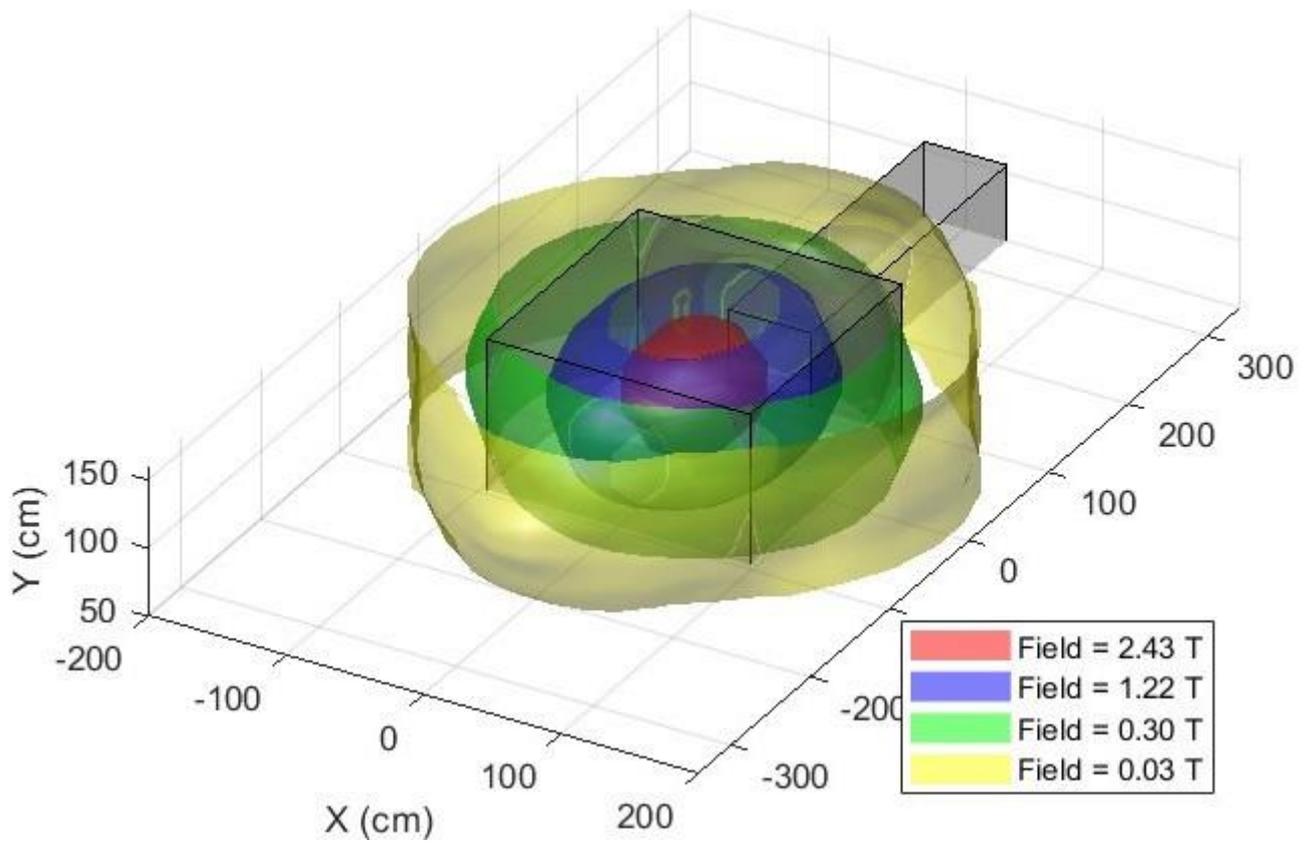